\documentclass[10pt]{article}
\usepackage[dvipdfm]{graphicx,color}
\title{Free-Energy Landscape of Kinesin by a Realistic Lattice Model}
\date{}

\author{Hiroo Kenzaki$^{1 2 3}$, and Macoto Kikuchi$^{2 1 4}$\\
$^1$ Department of Physics, Osaka University, Toyonaka 560-0043, Japan\\
$^2$ Cybermedia Center, Osaka University, Toyonaka 560-0043, Japan\\
$^3$ Core Research for Evolutional Science and Technology,\\
Japan Science and Technology Agency, Nagoya, Japan\\
$^4$ Core Research for Evolutional Science and Technology,\\
Japan Science and Technology Agency, Suita, Japan}

\begin{document}
\maketitle


{\bf Present address}: Department of Computational Science and Engineering, Nagoya University, Nagoya 464-8603, Japan


{\bf E-mail address}: kenzaki@tbp.cse.nagoya-u.ac.jp

\begin{abstract}

Structural fluctuations in the thermal equilibrium of the kinesin motor domain
are studied using a lattice protein model with G\={o} interactions.
By means of the multi-self-overlap ensemble (MSOE) Monte Carlo method
and the principal component analysis (PCA), the free-energy landscape is obtained.
It is shown that kinesins have two subdomains that
exhibit partial folding/unfolding at functionally important regions:
one is located around the nucleotide binding site
and the other includes the main microtubule binding site.
These subdomains are consistent with structural variability
that was reported recently based on experimentally-obtained structures.
On the other hand, such large structural fluctuations
have not been captured by B-factor or normal mode analyses.
Thus, they are beyond the elastic regime, and it is essential to take into account chain connectivity for studying the function of kinesins. 
\end{abstract}

\section*{INTRODUCTION}

Biomolecular motors are proteins that exhibit biolocomotion
by converting chemical energy into mechanical energy
\cite{mehta99,vale00,schliwa03}.
A detailed mechanism of this energy conversion has been a long-standing problem.
Unlike macroscopic motors, biomolecular motors operate under a strong influence of thermal fluctuations,
since the chemical energy provided by ATP hydrolysis is of the same order as that of the thermal fluctuations. 

Kinesin, which is responsible for intracellular transport and cell division
\cite{hirokawa98},
moves stepwise in one direction along the microtubule with one ATP hydrolysis per step
\cite{svoboda93,schnitzer00,taniguchi05}.
Kinesin has attracted considerable attention and was studied intensively in recent years.
Conventional kinesin functions as a dimer;
it is widely accepted that kinesin walks hand-over-hand on a microtubule
\cite{yildiz04}.
The gliding movement of kinesin is believed to be realized
by controlling the binding with and dissociation from the microtubule
\cite{cross04}.
Kinesin binds tightly to the microtubule in the nucleotide-free and ATP states,
whereas it detaches from the microtubule in the ADP state
\cite{romberg93}.

The structures of kinesin and myosin share highly conserved regions known as switch I and switch II, named after their counterparts in G proteins \cite{vale96};
G proteins are signaling proteins that bind to GTP instead of ATP.
The switch regions of the G proteins move during nucleotide hydrolysis
and thereby control the binding with and detaching from a target protein
\cite{vetter01}.
As for kinesin, both the switch I and switch II regions have different conformations among different X-ray structures
\cite{sack99,kull02,nitta04};
further, they show microtubule-dependent structural changes
\cite{rice99}.
This strongly suggests that kinesin also controls its affinity with the rail protein by structural transformations of the switch regions. 

It is commonly considered that functionally important fluctuations of a protein can be read from the B-factor of the X-ray crystal structure.
For kinesin, however, it was reported that
a large B-factor does not correspond to a (supposedly) functionally important fluctuation \cite{sack99}.
In particular, it was pointed out that the fluctuations of the switch regions are largely underevaluated.
Moreover, a normal mode analysis based on the elastic-network model
has revealed that small fluctuations in the elastic regime are insufficient
to capture the conformational change of kinesin
in contrast to the other molecular motors
\cite{zheng03}.
These results indicate that larger-scale structural fluctuations beyond the elastic regime are relevant to the function of kinesins.
In order to investigate such large structural fluctuations,
we follow a methodology of folding study in this work.

The energy landscape theory of protein folding has been accepted widely in the last decade;
this theory states that proteins have a funnel-like energy landscape toward the native structure
\cite{bryngelson95,onuchic97}.
This new view of the protein folding has resulted in a revival of the G\={o} model
\cite{taketomi75,go83},
which was originally introduced for explaining the mechanism of two-state folding transitions of small globular proteins.
Recent studies of protein folding have shown that G\={o}-like models
(namely, models constructed with the same design as the G\={o} model)
are applicable far beyond the original scope;
they can describe the folding of some proteins with intermediate states
\cite{zhou97,clementi00,koga01,karanicolas03,levy04,das05,kenzaki06,kenzaki06b}.
In this work, we examine the large structural fluctuations of kinesins by applying a G\={o}-like model for calculating a free-energy landscape.
We use a realistic lattice protein model
\cite{kenzaki06,kenzaki06b},
because their relatively small conformation spaces are fairly favorable for numerical simulations. 

\section*{MODEL AND METHODS}

In the lattice model of its simplest form,
each amino acid residue is expressed as a bead that is allowed to move on a simple cubic lattice,
and the polypeptide chain is represented as a string of connected beads with a unit bond length
\cite{sali94,dill95}.
If a pair of unconnected beads occupies neighboring sites,
the pair is called to form a contact and the interaction is applied to it.
Although such a lattice model can reproduce various universal properties of the protein folding,
it is too simple to represent the chain geometry of real proteins;
thus, we need to introduce a lattice model of a higher resolution for the present purpose.

Some realistic lattice models that can represent flexible protein structures have been proposed
\cite{kolinski96,kolinski04,hao96}.
In the present work, we use a 210-211 hybrid lattice model in which C$^{\alpha}$  atoms of amino acids are again located on a simple cubic lattice;
however, consecutive atoms are connected by vectors of the type (2,1,0) or (2,1,1) and all their possible permutations
\cite{kenzaki06,kenzaki06b}.
The length of (1,0,0) vector is set to $1.62$ \AA,
and the length of a C$^{\alpha}-$C$^{\alpha}$ bond is then $3.63$ \AA{} or $3.97$ \AA.
In order to express the excluded volume effect,
we consider that each amino acid occupies seven lattice sites,
that is, a center site at the bead position and its nearest neighbors.

We introduce G\={o} interactions between amino acids,
which act only on pairs of amino acids that form native contacts.
We call such pair a "native pair" from now on.
The X-ray crystal structure of the kinesin motor domain of {\it Homo sapiens} (PDB ID code 1bg2) is used as a reference of the native structure (Figure 1(a)) \cite{kull96}.
This crystal structure is of the ADP-binding state
(note that the ADP molecule is not included explicitly in the following simulations).
The native structure of the lattice protein model is defined by fitting to the C$^{\alpha}$ trace of this crystal structure.
The accuracy of the fitting in terms of the root mean square deviation (rmsd) is $0.92$ \AA (Figure 1(b)).
The G\={o} potential between the $i$ th and $j$ th amino acids is defined as follows:

\begin{eqnarray}
V_{g\bar{o}} & = & - \sum_{j-i>2} \varepsilon_{i,j} C_{i,j}
                     \Delta (r_{i,j}, r_{i,j}^{nat}),\\
\Delta(x,y) & = & \left\{
\begin{array}{lc}
1 & |x^2 - y^2| \le W\\
0, & \textrm{otherwise}
\end{array} \right.
\end{eqnarray}

\noindent
where $r_{i,j}$ is the distance between C$^{\alpha}$ atoms, 
$r_{i,j}^{nat}$ is their native distance,
$\varepsilon_{i,j}$ is the interaction strength between residue $i$ and $j$,
and $C_{i,j}$ is the constant that equals $1$ or $0$ 
depending on whether the pair is a native pair or not.
$W$ is the width of the G\={o} potential.

We add a local potential to maintain a protein-like local structure.
Here, we introduce a harmonic potential between the $i$ th and ($i+2$) th residues to bias the bond angles toward the native structure. 
The local potential is

\begin{eqnarray}
V_{local} & = & \frac{K_{b}}{2}
\sum_{i=1} (r_{i,i+2} - r_{i,i+2}^{nat})^2
\end{eqnarray}

\noindent
where $K_{b}$ is the strength of the local interaction. 
We use $W = 2$ and $K_b = 1$.
The interaction strength $\varepsilon_{i,j}$ takes
either the uniform value $\varepsilon_{i,j} = 1$ or
normalized Miyazawa-Jernigan (MJ) contact-energy parameters
\cite{miyazawa96}.

We compute the density of states and several physical quantities in thermal equilibrium
by using a Monte Carlo scheme called multi-self-overlap ensemble (MSOE) Monte Carlo method \cite{iba98,chikenji99}.
In this method, the self-avoiding condition is systematically weakened,
and a flat histogram both in energy and degree of the excluded volume is obtained by a bivariate extension of the multicanonical ensemble Monte Carlo method.
By calculating the physical quantities only for the self-avoiding conformations,
correct canonical averages are obtained.
The MSOE method gives a high performance, especially for long proteins,
in comparison with the standard multicanonical ensemble for this realistic lattice model. 
It should be noted that both lattice model and multicanonical methods are not suitable for studying dynamics of protein.
In the present work we focus on equilibrium properties, the free-energy landscape in particular,
which can be treated by the lattice model as long as the conformations are realistically expressed.

\section*{RESULTS}

Simulations are prepared for the following three models:
(i) C$^{\alpha}$($7.5$): A pair of amino acids is considered to be a native pair
if the distance between their C$^{\alpha}$ atoms is less than $7.5$ \AA{} in the X-ray crystal structure.
(ii) All($4.5$): A pair of amino acids is considered to be a native pair
if the minimum distance between their heavy atoms is less than $4.5$ \AA{} in the X-ray crystal structure.
(iii) All($4.5$)/MJ: Definition of the native pair is same as All($4.5$),
but heterogeneous interactions based on normalized MJ contact-energy parameters \cite{miyazawa96} is used as the G\={o} interaction. 
The density of state of kinesin across a wide energy range
including both the native state and unfolded state are calculated by MSOE.
Figure 2(a) shows the entropy of the kinesin motor domain.
The figure represents the folding funnel if rotated by $90$ degrees.

\subsection*{Free-Energy Landscape and Fluctuation of C$^{\alpha}$($7.5$) Model}

We first focus on the C$^{\alpha}$($7.5$) model.
The energy shown in Figure 2(b) (solid line) jumps at two temperatures.
Both the jumps correspond to cooperative transitions.
The gyration radius significantly changes at the higher transition point,
and slightly at the lower transition point (Figure 2(c)).
As shown below, the hydrophobic core is broken and kinesin unfolds completely above the higher transition point.
Thus, the higher transition is the main folding/unfolding transition
and is considered to be irrelevant to the stepping motion.
On the other hand, folding/unfolding of local structures takes place at the lower transition point with the hydrophobic core maintained folded.

In order to examine each transition in detail,
we carry out the principal component analysis (PCA) for contact formation of the native pairs.
The variable $x_k$ is introduced,
where the index $k$ runs all the native pairs;
$x_k = 1$ if $k$-th native pair is in contact and 0 otherwise.
The variance-covariance matrix 
\begin{equation}
M_{kl} = \langle x_kx_l \rangle - \langle x_k \rangle \langle x_l\rangle
\end{equation}
is calculated, where $\langle \rangle$ indicates thermal average.
Then $M_{kl}$ is diagonalized to yield the eigenvalues and eigenvectors.
The eigenvectors are called the principal components (PC);
PC expresses which contacts are formed/unformed cooperatively, and PC with a large eigenvalue represents an important structural fluctuation. 
We calculate the average value of amplitudes in PC of all the contacts belonging to each residue.
Residues are regarded to constitute a fluctuation unit, if they have relatively large averages in the same PC.

At the higher transition point, only one eigenvalue is large
and the fluctuation unit is located at the hydrophobic core [central anti-parallel $\beta$-sheet (Figure 1(a))].
The eigenvalue distribution at the lower transition point is shown in Figure 3(a).
There are two large eigenvalues.
The fluctuation unit of the first eigenvector (PC1) is composed of
helix 4, loop 12, helix 5, helix 6, and strands 5a/b (insertion of strand 5),
and that of the second eigenvector (PC2) is composed of helix 3 and helix 2 [Figure 3(b)]
We found that structural fluctuations of these fluctuation units actually are partial folding and unfolding,
because the components in the principal eigenvectors have the same sign.
The former region overlaps with the switch II region and the latter with the switch I region.
Hence, we call these regions SWII subdomain and SWI subdomain, respectively.

Figure 3(c) shows the free-energy landscape in two-dimensional space
spanned by the first two principal components PC1 and PC2 at the transition point.
Four states are distinguishable as four minima in the free-energy landscape.
They are classified according to the formation of SWI and SWII:
both SWI and SWII are unfolded (UU state);
only SWI is formed (FU state);
only SWII is formed (UF state);
both the subdomains are formed (FF state).
The two subdomains behave as mutually independent folding units because they belong to different eigenvectors.
In other words, partial folding/unfolding transitions of
FF $\leftrightarrow$ FU $\leftrightarrow$ UU and
FF $\leftrightarrow$ UF $\leftrightarrow$ UU occur independently.
It should be noted that the FU $\leftrightarrow$ UF transition is not realized
because of the free-energy barrier.

If these fluctuation units just reflect intrinsic properties of the lower transition point and do not have any relevance for other temperature, they may not be of interest from the viewpoint of function of kinesin.
In order to demonstrate that this is not the case and the fluctuation units are rather robust, 
we show the energy dependence of the mean contact probability of the native contacts for each residue in Figure 3(d).
Five states are distinguishable,
each of which corresponds to a different energy range.
The highest energy range corresponds to the completely unfolded state,
so that it is not relevant to the lower transition.
The $\beta$-sheet hydrophobic core is formed in the other four states.
They actually correspond to UU, FU, UF and FF states described above
and appear in Figure 3(d) in this order from a higher energy to a lower energy.
Thus, we can safely say that SWI and SWII subdomains behave as two independent fluctuation units as long as the hydrophobic core is formed.

\subsection*{Units of Fluctuation Are Robust}

Let us now turn to the All($4.5$) and All($4.5$)/MJ models.
As is seen in Figure 2(b), only a single jump is observed in the energy.
The protein is completely unfolded at temperatures above this jump.
Hence, in these cases, the folding/unfolding of local structures are not separated from that of the hydrophobic core with respect to the temperature.
Thus, PCA is not suitable to describe partial folding/unfolding.
Nevertheless, we can discuss the folding unit from the energy dependence of the mean contact probability of the native contacts.
As is seen in Figures 4(a) and 4(b),
five states - unfold, UU, UF, FU, and FF - are again distinguishable.
In other words, they are separated in energy similar to the C$^{\alpha}$($7.5$) models.
Thus, two subdomains - SWI and SWII - are also included as the folding units in both the All($4.5$) and All($4.5$)/MJ models.
The structure of the boundary of the subdomains is slightly different;
strand 1 is included in the hydrophobic core in the C$^{\alpha}$($7.5$) model,
but in the SWII subdomain in the All($4.5$ ) and All($4.5$)/MJ models.
Strand 5a/b is included in the SWII subdomain in the C$^{\alpha}$($7.5$) model,
but not in the All($4.5$) and All($4.5$)/MJ models.
In brief, although the number of transitions is different for different models,
the overall structure of folding units (subdomains) is robust.

\section*{DISCUSSION}

The simulation result indicate that
the kinesin motor domain has two subdomains other than the hydrophobic core:
SWI and SWII, which overlap with the switch I and switch II regions, respectively.
These two subdomains exhibit structural fluctuations,
which actually are partial folding/unfolding, and nearly independent when the nucleotide is absent.
The largest fluctuation is localized at the SWII subdomain. 
A question may occur whether these subdomains are relevant to the real kinesin protein or not.
Quite recently, Grant et al. studied
structural variability of kinesin using 37 available structures in the protein data bank (PDB) \cite{grant07}.
The PCA revealed that there are two significant principal components. 
The first principal component describes the concerted displacement located at helix 4, loop 12, helix 5, and loop 13.
This subdomain corresponds to the central part of the SWII subdomain we found.
They suggested that the second principal component is localized at helix 3 and the vicinity of loop 6 and loop 10.
Seeing their result, we found that the component has a large value at helix 2, which actually is located in the vicinity of loop 6.
Although they did not call this region as subdomain explicitly, the region and the SWI subdomain overlap in large part with each other.
Thus, our simulation result is consistent to their analysis.
This indicates that the SWI and SWII subdomains successfully captured the functionally important fluctuations.

These two subdomains are considered to play important roles in the stepping motion,
based on the following experimental facts:
(i) Helix 4, loop 12, and helix 5 belonging to the SWII subdomain
are known to undergo a large conformational change during the ATP hydrolysis cycle
\cite{nitta04};
this region also includes the main microtubule binding site
\cite{woehlke97,hoenger00}.
Moreover, the SWII subdomain interacts directly with the neck linker,
which connects the motor domain and neck region (the neck linker is not included in our simulation),
because the neck linker is adjacent in sequence to helix 6 (included in the SWII subdomain).
(ii) Helix 3 in the SWI subdomain also undergoes a large conformational change during the ATP hydrolysis cycle \cite{naber03}.
In addition, judging from the number of included residues,
we may safely assume that a time scale of the conformational changes of these two subdomains of real kinesin
is approximately the same order as that of the stepping motion ($\sim$ msec).
In contrast, it is unlikely that the hydrophobic core unfolds on the stepping motion,
because the time scale of the core unfolding should be much longer ($\sim$ sec or longer).
The present results suggest that conformational changes of these functionally important regions are realized by a partial unfolding-refolding.
We showed that SWI and SWII behaves as independent folding unit in the absence of nucleotide.
Binding to the nucleotide may induce coupling between these two folding units, which will be left for future study.

The subdomain structure is largely determined
by the topology of the native structure
because three variations of coarse-grained G\={o}-like models
show similar subdomain structures.
It has been pointed out that the cooperativity
of folding transition depends on the detail of the interaction
within the framework of the G\={o}-like model,
such as inclusion of many-body interaction \cite{chan04},
interaction range \cite{kenzaki06}, and so on.
However, as we found in previous study
subdomain structure is rather insensitive to such detail \cite{kenzaki06}.

We note that the large fluctuations of these subdomains are totally different from the small fluctuations observed by,
for example, the B-factor in the X-ray crystal structure and elastic network model.
In fact, the B-factor of the subdomains is not particularly larger than that of other parts of the motor domain.
Thus, the large fluctuations described by the present model are beyond the elastic regime and could not be captured by the B-factor or elastic network model.
It is essential to take into account the chain connectivity explicitly for studying the function of kinesins.
Myosin motor domain is more than twice the size of kinesins,
but their nucleotide binding regions and switch I/II regions have similar topology to those of kinesins.
This indicates that the basic properties of the free-energy landscape of myosin may also be similar.

Before closing the discussion,
some comments should be made on related simulations.
All-atom molecular dynamics simulations of the kinesin motor domain have been performed
\cite{wriggers98,minehardt01}.
Wriggers et al. captured the nucleotide-dependent small fluctuations of the switch regions \cite{wriggers98},
and Minehardt et al. simulated conformational change of the switch I region \cite{minehardt01}.
Although, these results are in part consistent with the present result,
these simulations are extremely short in time (order of nanosecond)
as compared to the time scale of the stepping motion (order of millisecond),
since the all atom simulation of a large protein as kinesin is still a hard task.
Recently, Hyeon and Onuchic performed the simulation
of the kinesin dimer bounded on the microtubule by the G\={o}-like model,
and investigated the allosteric transition regulated
by the inter-monomer strain through the neck-linker \cite{hyeon07}.
In contrast, 
we focused on the structural fluctuations of a kinesin monomer in thermal equilibrium.
We found that kinesin has two subdomains, which give rise to the conformational fluctuation of the functionally important regions.
This result is supported by the recent analysis on experimental data \cite{grant07}.
The coarse-grained G\={o}-like model made it possible to describe
the large structural fluctuation that is relevant to the function.

\section*{ACKNOWLEDGMENTS}
We thank G. Chikenji, M. Sasai and F. Takagi for fruitful discussions.
The present work is partially supported by
the IT-program of Ministry of Education, Culture, Sports, Science and Technology,
the 21st Century COE program named "Towards a new basic science: depth and synthesis.",
and Grant-in-Aid for Scientific Research (C) (17540383) from Japan Society for the Promotion of Science.

\clearpage
\begin{figure}
\includegraphics{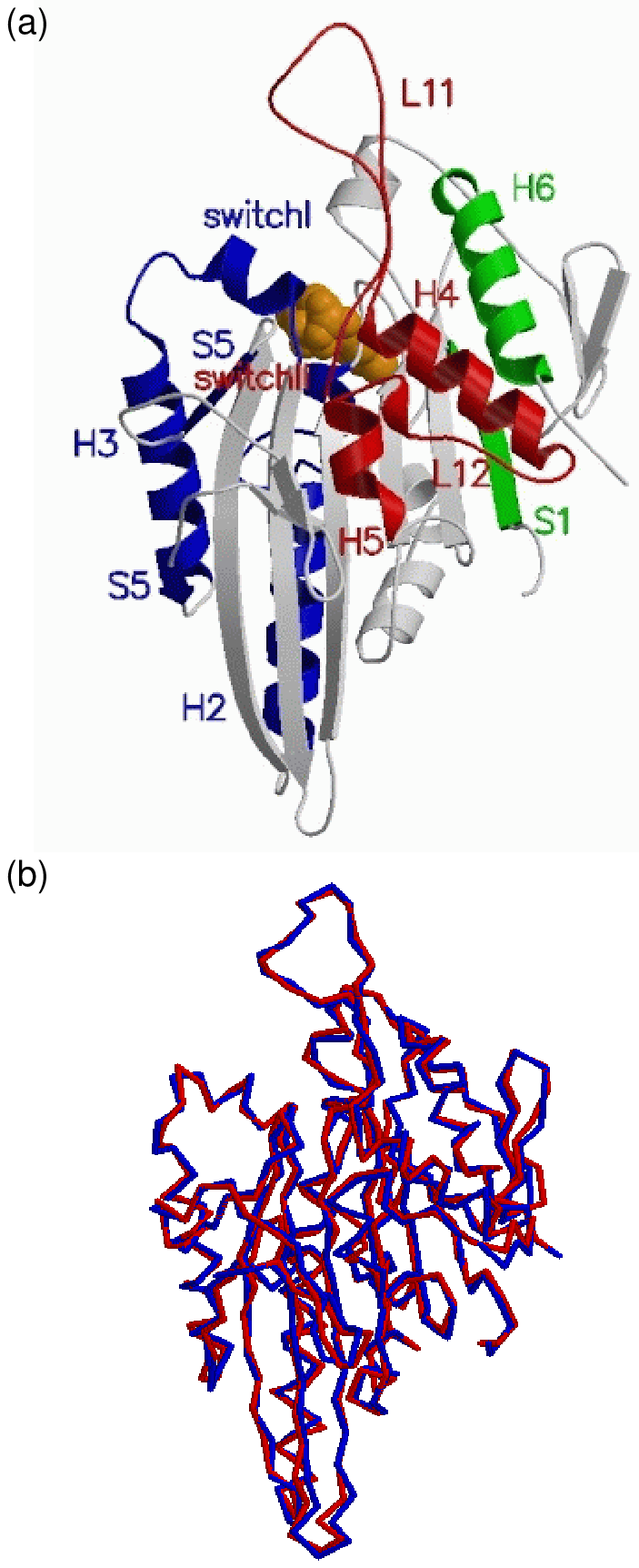}
\caption{
(a) The X-ray structure of kinesin motor domain (PDB ID code 1bg2).
It consists of a central antiparallel $\beta$-sheet of eight strands,
sandwiched between three $\alpha$-helices on either side.
Orange: ADP, blue: helix 2 (91-122), strand 5 (141-144, 171-173),
helix 3 (176-189) and switch I (198-203),
red: switch II (231-236), loop 11 (238-254), helix 4 (257-269),
loop 12 (272-280), and helix 5 (281-292),
green: strand 1 (9-15), and helix 6 (306-320).
This figure was prepared using the programs
MOLSCRIPT \cite{kraulis91} and Raster3D \cite{merritt97}.
(b) The superposition of native structure of the lattice protein model of kinesin (blue)
and the C$^{\alpha}$ trace of the X-ray crystal structure (red).
Accuracy of the fitting in terms of
the root mean square deviation (rmsd) is $0.92$ \AA.}
\end{figure}

\begin{figure}
\includegraphics{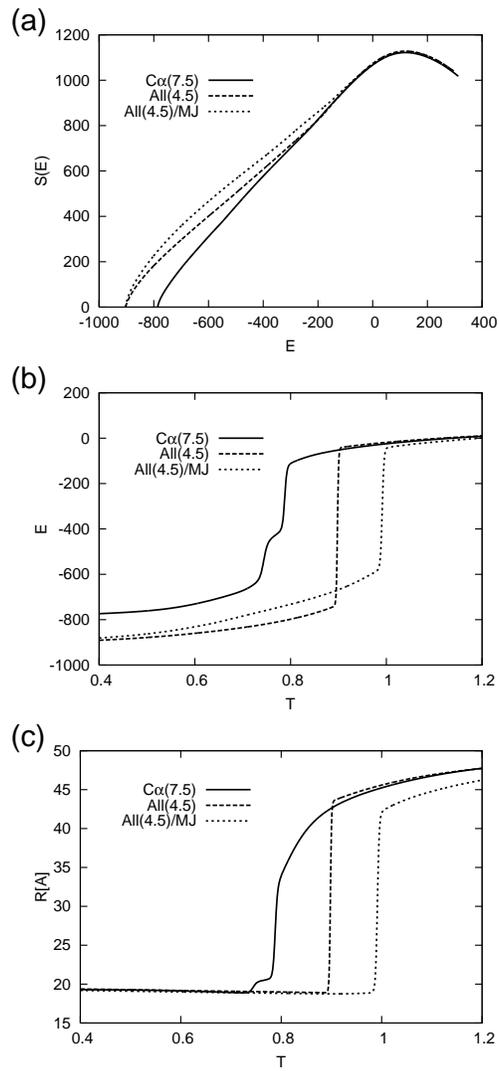}
\caption{
(a) Relative entropy $S(E)$ of kinesin as a function of energy $E$.
(b) The average energy $E$ as a function of temperature $T$.
(c) The radius of gyration $R$ as a function of temperature $T$.}
\end{figure}

\begin{figure}
\includegraphics{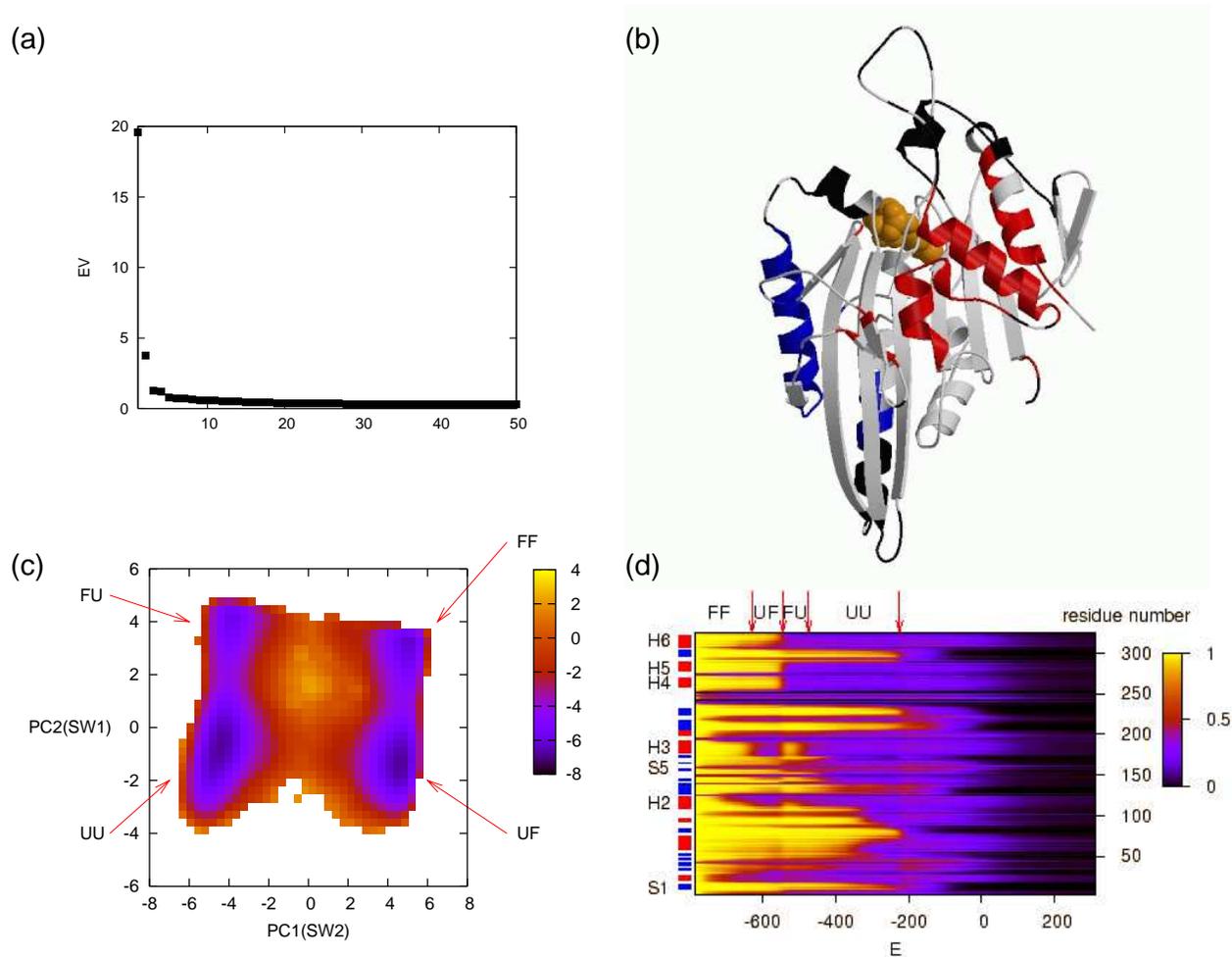}
\caption{
Free-energy landscape and structural fluctuation of C$^{\alpha}$(7.5) model.
(a) Eigenvalue distribution of PCA at the lower transition point.
(b) The first two principal components PC1 and PC2 at the lower transition point are shown.
The residues belonging to the fluctuation unit of PC1 are shown in red
and those of PC2 are blue.
Black indicates unstable residues with low contact formation probabilities.
ADP (not included in the simulation) is orange.
(c) Free-energy landscape at the lower transition point
in two dimensional space spanned by the first two principal components PC1 and PC2.
(d) Energy dependence of the mean contact probability of the native contact
for each residue.}
\end{figure}

\begin{figure}
\includegraphics{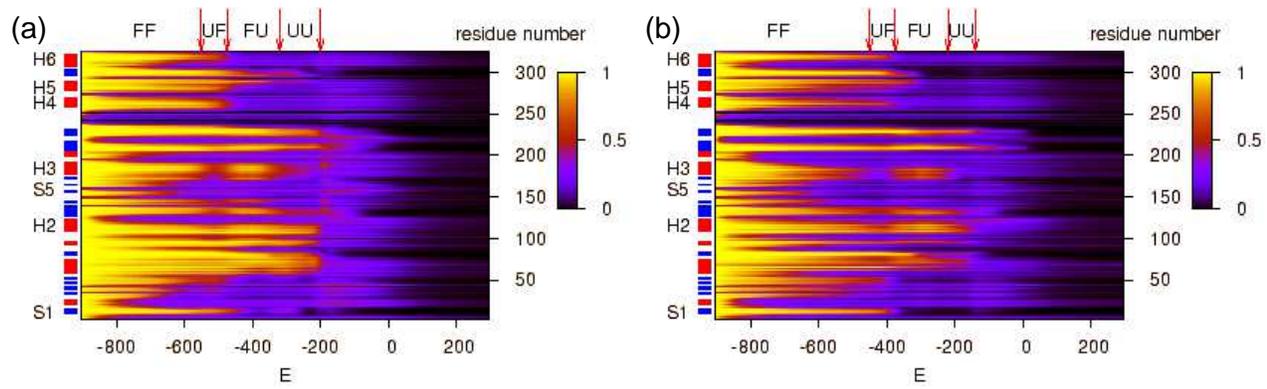}
\caption{
Energy dependence of the mean contact probability of the native contact
for each residue for (a) All($4.5$) and (b) All($4.5$)/MJ.}
\end{figure}

\end{document}